\documentclass{aastex631}
\usepackage{graphicx}

\shorttitle{Dust Emission as a Function of Age in M33}
\shortauthors{Mallory et al.}

\newcommand{\hii}{\hbox{H\,{\sc ii}}}
\graphicspath{{./}{figures/}}

\begin{document}

\title{Dust Emission as a Function of Stellar Population Age in the Nearby Galaxy M33}

\author[0000-0002-2898-1416]{Kate Mallory}
\affiliation{Department of Astronomy, University of Massachusetts Amherst, 710 North Pleasant Street, Amherst, MA 01003, USA}

\author[0000-0002-5189-8004]{Daniela Calzetti}
\affiliation{Department of Astronomy, University of Massachusetts Amherst, 710 North Pleasant Street, Amherst, MA 01003, USA}

\author[0000-0001-8078-3428]{Zesen Lin}
\affiliation{CAS Key Laboratory for Research in Galaxies and Cosmology, Department of Astronomy, University of Science and Technology of China, Hefei 230026, People's Republic of China}
\affiliation{School of Astronomy and Space Science, University of Science and Technology of China, Hefei 230026, People's Republic of China}

\begin{abstract}
Dust emission at 8~$\mu$m has been extensively calibrated as an indicator of current star formation rate for galaxies and $\sim$kpc-size regions within galaxies. Yet, the exact link between the 8~$\mu$m emission and the young stellar populations in galaxies is still under question, as dust grains can be stochastically heated also by older field stars. In order to investigate this link, we have combined mid-infrared images from the {\sl Spitzer Space Telescope} with a published star cluster candidates catalog for the Local Group galaxy M33. M33 is sufficiently close that the {\sl Spitzer's} 8~$\mu$m images resolve  individual regions of star formation. Star clusters represent almost--single--age stellar populations, which are significantly easier to model than more complex mixtures of stars. We find a decrease in the 8~$\mu$m luminosity per unit stellar mass as a function of age of the star clusters, with a large scatter that is consistent with varying fractions of stellar light absorbed by dust. The decrease and scatter both confirm findings based on more distant galaxies and are well described by simple models for the dust emission of a young stellar population. We conclude that the dust emission at 8 $\mu$m depends sensitively on the age of the stellar population, out to at least the oldest age analyzed here, $\sim$400~Myr. This dependence complicates the use of the 8~$\mu$m emission as a star formation rate indicator, at least for small galactic regions and individual star forming regions. By leveraging the {\sl Spitzer} legacy, this investigation paves the way for future explorations with the {\sl James Webb Space Telescope}.
\end{abstract}

\keywords{Interstellar Dust: Polycyclic Aromatic Hydrocarbons; Disk Galaxies: M33 (Triangulum Galaxy); Young Star clusters}

\section{Introduction} \label{sec:intro}

Star formation occurs in dusty environments, and most of the ultraviolet (UV) and optical light emitted by the young stars is absorbed by dust and re-emitted in the mid- and far-infrared, at $\lambda>$3~$\mu$m. Thus infrared emission can effectively trace the star formation in galaxies and regions within galaxies. For this reason, much effort has been expended over the past few decades to calibrate the infrared emission as a star formation rate (SFR) tracer \citep[see review by][]{Kennicutt}.

The mid-infrared ($\lambda\sim$3--30~$\mu$m) emission is particularly attractive for tracing recent star formation, as it can be targeted by infrared facilities out to moderate redshifts. Samples of galaxies out to z$\approx$2 were secured by observations in the restframe mid-infrared with the {\sl Spitzer Space Telescope} \citep[{\sl Spitzer} henceforth,][]{Werner+2004} and the {\sl Herschel Space Observatory} \citep[e.g.,][]{Reddy+2006, Reddy+2010, Elbaz+2011, Shivaei, Shipley, Shivaei2017}; higher redshift populations, using the bluest mid-infrared features, can potentially be secured with the recently launched James Webb Space Telescope due to its higher sensitivity relative to the other telescopes. 

The mid-infrared region is dominated by emission from Polycyclic Aromatic Hydrocarbons (PAHs) and small dust grains stochastically heated by a wide range of photon energies in the galaxy's radiation field \citep{Leger, DraineLi, Tielens2008}. The PAH molecules, which produce $\sim$5\%--10\% of the total infrared emission \citep{Smith+2007}, are excited by UV photons in the wavelength range $\sim$0.0912--0.2~$\mu$m and emit in a series of features  between 3~$\mu$m and $\sim$20~$\mu$m. These include the strong $\sim$7.7~$\mu$m emission feature which dominates the band targeted by the {\sl Spitzer}/IRAC~8~$\mu$m array \citep{Fazio} 
and has been extensively calibrated as a SFR tracer \citep[e.g.,][]{Wu+2005, Reddy+2006, Kennicutt+2009, Shipley}.  As discussed in \citet{Shipley}, calibrations from different authors are generally consistent with each other within $\sim$0.2~dex, when using samples of metal-rich galaxies dominated by recent star formation (SFRs$>$a few M$_{\odot}$~yr$^{-1}$). 

PAHs, however, are also fragile molecules that are destroyed in the high-energy environments of massive star formation and AGNs,  requiring shielding from large dust grains to survive \citep{Madden+2006, Gordon+2008, Lebouteiller+2011, Binder+2018}. For this reason, PAH emission is mainly found in the Photodissociation Regions surrounding star forming regions and in the diffuse interstellar medium \citep{Povich+2007, Bendo+2008, Relano+2009, Tielens2013}. In local galaxies, between 30\% and 80\% of the emission at 8~$\mu$m is diffuse, found outside of regions that actively form stars \citep{Boselli+2004, Crocker+2013, Calapa+2014, Lu+2014}. 

The complexity of the response of the PAH features to heating and excitation is compounded by their strong sensitivity to the metal abundance of the interstellar medium \citep[e.g.,][]{Engelbracht+2008}. Their intensity becomes negligible below oxygen abundances $\sim$0.2 solar \citep{Aniano+2020}, which has been interpreted as due to the formation mechanism  \citep{Sandstrom+2012} or processing of the carriers by the environment \citep{Madden+2006, Gordon+2008}. These dependencies ultimately have complicated a robust calibration of the 8~$\mu$m emission as a SFR tracer \citep{Calzetti+2007}.

\begin{figure}
\plotone{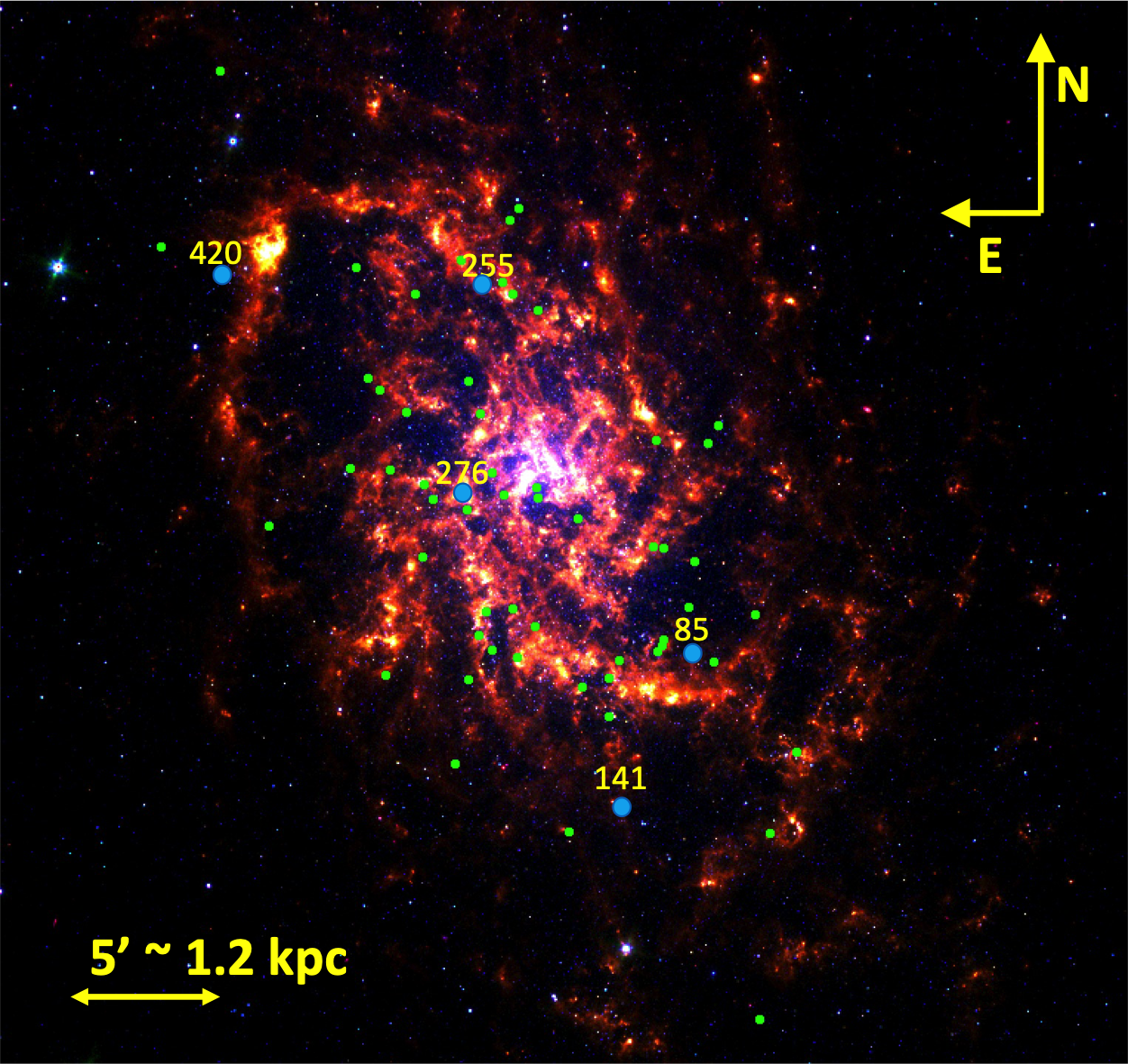}
\caption{A three-color image of the galaxy M33, at a distance of $\sim$ 850 kpc, obtained by combining the 4.5 $\mu$m (blue), 5.8 $\mu$m (green), and 8.0 $\mu$m (red) bands from the {\em Spitzer}/IRAC instrument. The scale of the image and orientation are shown on the figure. The 63 star clusters from the compilation of \citet{Sarajedini} used in this study are marked with green dots on the image. The five star clusters shown in detail in Figure~\ref{fig:Example} are marked with larger cyan dots and their IDs from \citet{Sarajedini} given on the galaxy image.}\label{fig:galaxy}
\end{figure}

One approach to disentangling the different effects acting on PAHs is to gauge the response of their emission to stellar populations with `controlled' parameters. Star clusters, by approximating single-age populations \citep{Wofford+2016}, offer convenient `controlled' templates for such analysis. They are abundant in galaxies, thus yielding the power of statistics, and cover a wide range of ages, from just formed systems to ancient sources almost as old as the Universe \citep[e.g.,][]{Adamo+2018}. The age of a stellar population is directly linked to its UV output \citep{Leitherer}, thus providing a tracer for this parameter.

Recently, \citet{Lin} investigated the 8~$\mu$m emission from 97 star clusters across 5 galaxies within 5~Mpc. The sample these authors use is from the LEGUS project \citep{Calzetti+2015}, which provided catalogs of ages, masses, and extinctions for $\sim$15,000 star clusters in 50 galaxies closer than $\sim$16~Mpc, using multiband {\sl Hubble Space Telescope} imaging \citep{Adamo+2017}. \citet{Lin} restricted their sample to nearby galaxies and relatively isolated star clusters to minimize superposition between sources measured with the lower-resolution {\sl Spitzer} and to mitigate contamination of the cluster emission from the surrounding stellar populations. Using their final  sample of 97 star clusters, the authors find a clear anti-correlation between the mass-normalized 8~$\mu$m luminosity of the star clusters and their age, indicating that the PAH emission directly responds to the presence of UV photons. The authors also show that the anti-correlation can be easily explained with simple models for the stellar populations combined with dust emission. They explain the significant scatter, $\sim$1~dex, in their data as combined effects of varying dust absorption fractions of the UV light and varying PAH abundances. Despite the careful selection, \citet{Lin} still cannot completely avoid potential effects of contamination of the observed 8~$\mu$m  emission from the stellar populations surrounding the star clusters. 

In this paper, we seek to expand upon the investigation of \citet{Lin} by analyzing the 8~$\mu$m  emission of the  star clusters in the Local Group galaxy M33. M33 (NGC598, Triangulum Galaxy) is located at a distance of 850~kpc  \citep{Ferrarese2000}, implying that the  2\arcsec\ full width at half maximum (FWHM) of the {\sl Spitzer}/IRAC 8~$\mu$m Point Spread Function subtends about  8~pc in spatial scale. M33 is an Sc spiral galaxy with a modest inclination of 52\arcdeg\ \citep[Figure~\ref{fig:galaxy};][]{Corbelli+2000}  
and significant star formation \citep[SFR$\simeq$0.5~M$_{\odot}$~yr$^{-1}$,][]{Verley2009}. Its proximity enables us to study the 8~$\mu$m emission from star clusters while drastically reducing the possibility of confusion from multiple sources along the line of  sight. We are also able to reach fainter 8~$\mu$m luminositities than was possible with the data used by \citet{Lin} and can better quantify the scatter at fixed age observed by these authors. 

The paper is organized as follows. Section~\ref{sec:s&c} presents the imaging data used in this work and the source of the star cluster catalog; Section~\ref{sec:processing} details the photometric measurements performed on the images; Section~\ref{sec:models} describes the models used for the comparison with the data; finally, in Section~\ref{sec:discussion} and Section~\ref{sec:conclusions} we discuss our results and our conclusions, respectively. 

\section{Data Sources and Characteristics} \label{sec:s&c}

The infrared images of M33 obtained with the {\sl Spitzer}/IRAC instrument were first presented by \citet{McQuinn+2007}. The images used in this study are those taken with the 3.6 $\mu$m and 8.0~$\mu$m filters. They were retrieved from the NASA Extragalactic Database (NED); the images are mosaics constructed by combining several frames, which provide a contiguous view of this large galaxy with a pixel scale of 0\farcs75/pixel and at a resolution (FWHM) of $\sim$1\farcs7 and  $\sim$2\arcsec\ for the 3.6~$\mu$m and 8.0~$\mu$m images, respectively \citep{Dale2009}. At the distance of M33 (previous section), these FWHMs correspond to physical scales of 7.0~pc  and 8.2~pc at 3.6~$\mu$m and 8.0~$\mu$m, respectively. The 3.6~$\mu$m image is used in this work exclusively to trace and remove the stellar contribution from the  8.0~$\mu$m emission, as described in the section  \ref{sec:processing}. The catalog of star clusters is taken from the compilation of \citet{Sarajedini}, which is described in the next section.

\section{Data Analysis} \label{sec:processing}

\citet{Sarajedini} classified sources in M33 into three categories: individual stars, star clusters, and unknown objects. Their compilation lists a total of 451 sources, including 255 confirmed clusters from {\sl Hubble Space Telescope} or high-resolution ground imaging. In addition to the classification, the authors provide the coordinates of the sources and, for many, optical photometry, ages and masses compiled from the literature \citep[see][for a list of references]{Sarajedini}. Briefly, the ages and masses are derived from fits with stellar population models of the 0.38--1~$\mu$m spectral energy distributions in eight medium-band filters \citep[e.g.,][]{Ma+2001}. 

We down-selected the \citet{Sarajedini} catalog as follows. We omitted all stars and several unknown sources which appeared to have stellar profiles from visual inspection of the optical images \footnote{also available from NED.}. Furthermore, we only selected sources that listed both ages and masses, and were younger that 400 Myr (i.e., $\log(\mathrm{age/yr}) < 8.6$). The latter constraint was imposed because 
older clusters are usually devoid of dust \citep{Whitmore2020}; for instance, \citet{Lin}'s sample only contains one cluster older than 400~Myr with detectable 8~$\mu$m emission at its location. An inspection of the stellar continuum-subtracted 8~$\mu$m image confirms that old clusters are generally coincident with non-detections. We finally removed all sources that are too faint to show emission at 8.0 $\mu$m and 3.6 $\mu$m. 
After trimming the list based on these criteria, we are left with 63 sources, which generally show a compact morphology at 8~$\mu$m. These sources are listed in Table~\ref{tab:Regions} together with the identification numbers, locations, ages, masses, and classification from \citet{Sarajedini}. Of the 63 sources, 13 are classified as `unknown' by those authors, while the rest are star clusters. We retain the 13 unknown sources since they show characteristics similar to those of the other star clusters. We note that \citet{Sarajedini} do not report uncertainties for their parameters. The locations of the 63 sources in M33 are indicated in Figure~\ref{fig:galaxy} with green and cyan dots. As expected for young star clusters, the sources are mainly located along the spiral arms, with 59/63 within 4~kpc of the center.

\begin{figure}
\plotone{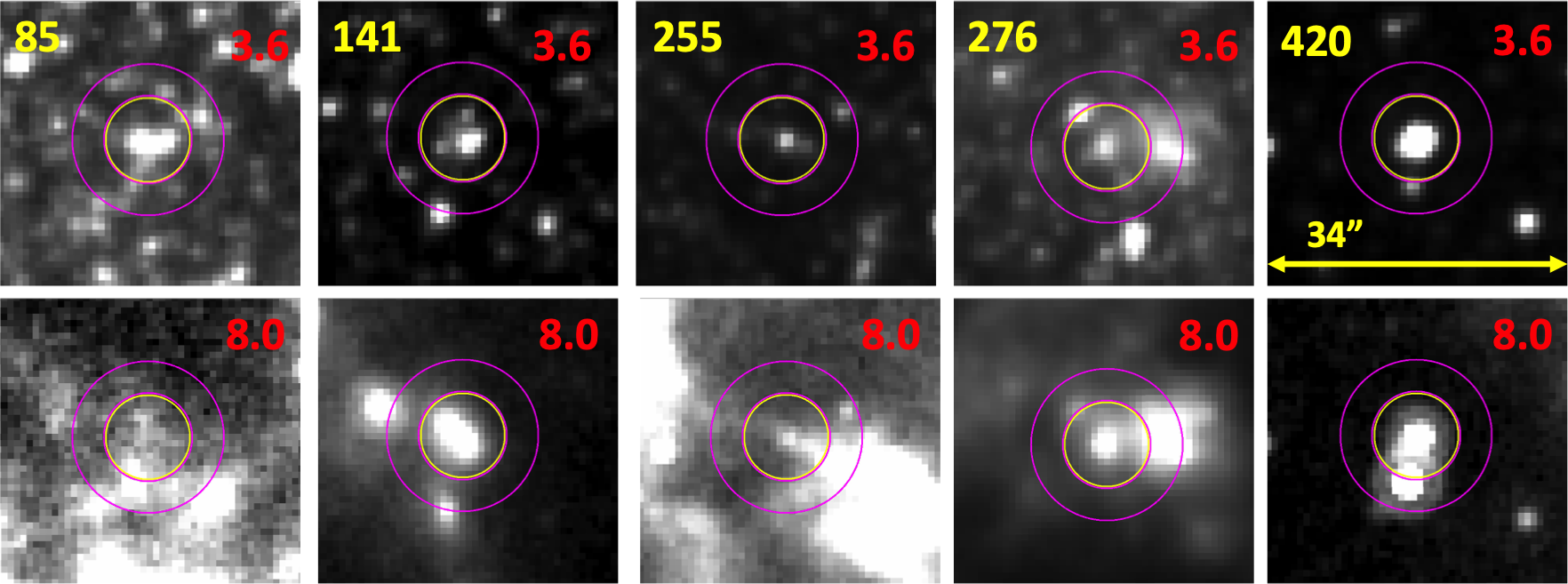}
\caption{Cutouts of the IRAC 3.6~$\mu$m (top row) and 8~$\mu$m (bottom row) images centered on the five sources marked in Figure~\ref{fig:galaxy}, showing the apertures (yellow circles) and the background annuli (magenta circles) used for the photometric measurements. The numbers given in the cutouts refer to the source IDs  listed in Table~\ref{tab:Regions}. \citet{Sarajedini} classify \# 141 and \# 276 as `unknowns', while the other three are confirmed star clusters. Each cutout is 34$^{\prime\prime}$ in size, as also indicated in the last panel to the right. The morphologies and behavior of the five clusters are representative of those of the overall sample of 63 sources.}
\label{fig:Example}
\end{figure}

Aperture photometry at both 3.6 $\mu$m and 8.0 $\mu$m was then performed with Astroconda, using a circular aperture centered on each candidate cluster. The aperture radii were chosen to be the same for both bands and for all clusters. We select an aperture radius of 6.67 pixels (5\farcs0, $\sim$21~pc) as an optimal compromise between the extent of the observed 8~$\mu$m emission and the need to avoid neighboring sources and inclusion of excess noise from the background. For measuring the background, we elect to use an annulus with inner radius of 7 pixels (5\farcs25) and a width of 5 pixels (3\farcs75). The goal of the annulus is to capture the local background values around each star cluster, with sufficient statistics to mitigate the noise in this measurements. In our case, the annulus includes about two times more pixels than the photometric aperture, thus reducing by sqrt(2) the noise level added by this component to the final measurement; the value of the background is calculated as the mode of the pixels' values, after implementing 10 iterations with  3-$\sigma$ clipping in order to remove the contribution of any bright source in the annulus. Examples of sources with the photometric apertures and sky annuli drawn on IRAC 3.6~$\mu$m and 8~$\mu$m image cutouts are given in  Figure~\ref{fig:Example}. The local background in the two images includes both the diffuse stellar light and diffuse dust emission (PAHs and stochastically heated dust) from the galaxy; by removing these, we concentrate on the stellar and dust emission associated with each star cluster. The 3.6~$\mu$m image may still contain hot dust emission heated by the young stars, but we expect this to only affect clusters younger than 10~Myr \citep{Whitmore2011} and to represent a small fraction of the background-subtracted 3.6~$\mu$m emission \citep{Querejeta2015}. For all practical purposes, the background-subtracted 3.6~$\mu$m photometry captures the stellar light from the star clusters.  The measured flux densities, $f_{3.6}$ and $f_{8.0}$, in the default units of MJy/sr of the {\sl Spitzer} images, are listed in Table~\ref{tab:Regions}. 

We apply aperture corrections to our measurements, in order to account for the flux in the wings of the cluster emission that is lost due to the finite size of the aperture used for the photometry.  While the 3.6~$\mu$m photometry is dominated by the photospheric emission of the stars in the cluster (see above), the 8~$\mu$m photometry is the  sum of contributions from stellar light and dust emission. Star clusters are compact sources, whose stellar distributions have typical sizes of $\sim$3~pc \citep{Ryon+2017, Brown+2021}. Conversely, the dust emission that surrounds the clusters is expected to be extended, as the UV photons that heat the dust can travel several tens of pc from the stars that produce them \citep{Relano+2009, Lawton+2010}. Because of this difference in behavior, aperture corrections for the dust emission at 8~$\mu$m are calculated after removing the stellar emission from this band.

At the distance of M~33, the typical size of star clusters translates to an angular size of $\sim$0.7$^{\prime\prime}$, much smaller than the IRAC $\sim$2$^{\prime\prime}$ Point Spread Function. Thus, for practical purposes, the stellar emission from our clusters can be considered originating from a point source. 
Table~4.8 of the IRAC Instrument Handbook \footnote{https://irsa.ipac.caltech.edu/data/SPITZER/docs/irac/iracinstrumenthandbook/} lists aperture corrections of $\sim$7\%  and $\sim$9\% for a point source measured in our default aperture radius (5\farcs0) at 3.6~$\mu$m  and 8~$\mu$m, respectively. We neglect, in what follows, the small 2\% difference  in correction between the two bands for the stellar emission.

\begin{figure}
\plotone{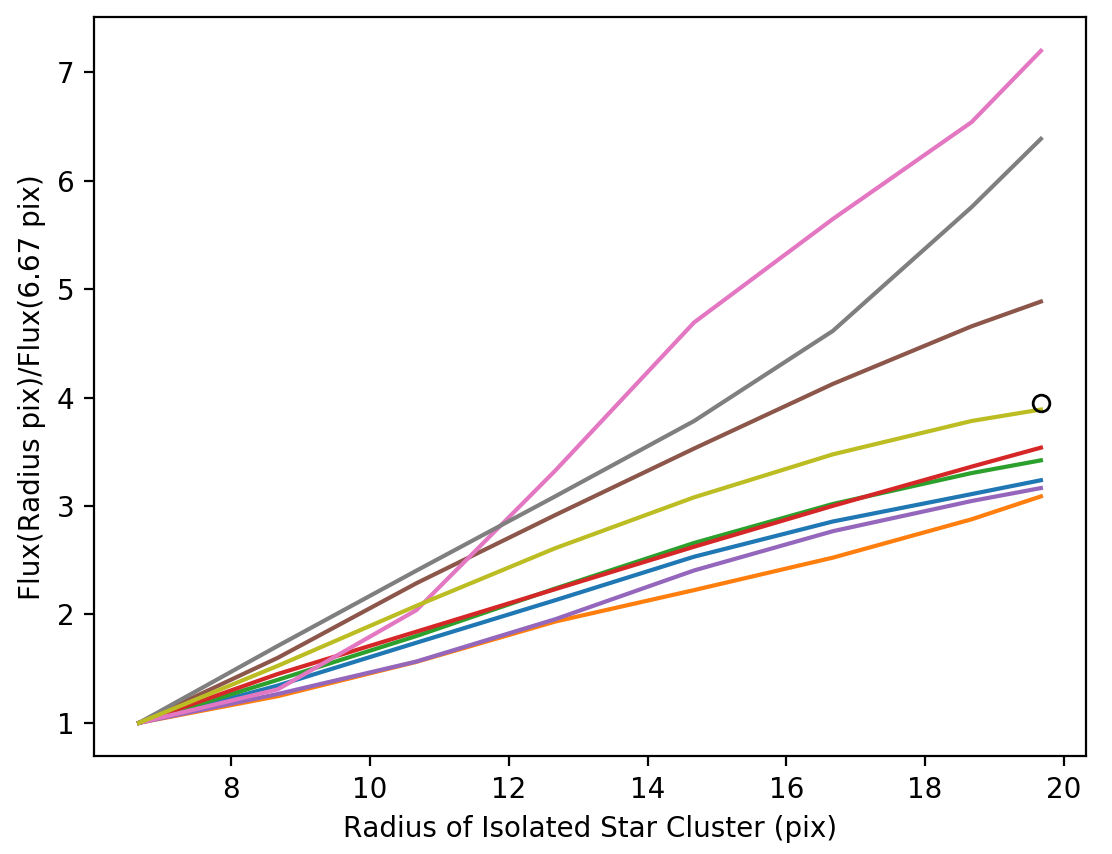}
\caption{Growth curves for 9 isolated clusters in the 8~$\mu$m image of  M33, for increasing radii from our default value of 6.67~pix  (5$^{\prime\prime}$) out to 20~pix (15$^{\prime\prime}$). The curves are derived from flux measurements at the indicated radii normalized to the flux in the default aperture. The open black  circle is the average of the curves at 20~pix, which we adopt as the 8~$\mu$m aperture correction for our photometric measurements.}\label{fig:NormFlux}
\end{figure}

The dust-only 8~$\mu$m emission is calculated as: $f_{8.0,\mathrm{ dust}}=(f_{8.0}-0.23 f_{3.6})$, following \citet{Helou2004}. As explained by these authors, the factor 0.23 applied to the 3.6~$\mu$m flux density rescales the stellar emission to its equivalent intensity at 8~$\mu$m, according to the Starburst99 stellar population models \citep{Leitherer}; the rescaled stellar component is then  subtracted from the observed  8~$\mu$m flux density to extract the dust-only 8~$\mu$m emission. 
To establish the aperture corrections for the dust emission, we perform photometry on 9 isolated clusters for increasing radii, from 6.67 pixels (our default aperture radius) to 20~pixels (15$^{\prime\prime}$, $\sim$60~pc). The maximum radius comfortably encloses a typical \hii\ region while ensuring that neighboring sources are not included in the measurements. 
The growth curves, normalized to the photometry in the default aperture (6.67 pixels) are shown in Figure~\ref{fig:NormFlux} for the dust--only 8~$\mu$m emission. We adopt the average of the growth curves at 20~pix as the mean aperture correction for our measurements; this correction increases our dust--only 8~$\mu$m photometric measurements by a factor of 3.95.  As a sanity check, we compare our aperture corrections with those derived by \citet{Lin}, taking into account the differences in galaxy distances  and aperture sizes between the two studies. We find that, when using closely matched {\sl physical} sizes for the apertures, the 8~$\mu$m aperture correction of \citet{Lin} is a factor $\sim$10\% smaller than our derived correction factor. We attribute the small difference to the higher spatial resolution of our study, which allows us to better sample the dynamical range of the dust-emitting regions. 

We convert the dust flux density to a luminosity at 8~$\mu$m, $L_{8.0}$, by multiplying the aperture--corrected flux density for the effective frequency of the 8~$\mu$m band (3.81$\times$10$^{13}$~Hz), using the distance of 850~kpc for M33. As our sources are extended, we include an additional correction factor 0.86 for light scattering into the IRAC instrument, which is appropriate for the largest aperture, 15$^{\prime\prime}$, used to calculate the curve of growth\footnote{section~8.2 of the IRAC Instrument Handbook: https://irsa.ipac.caltech.edu/data/SPITZER/docs/irac/iracinstrumenthandbook/}. The list of luminosities, corrected as described so far, is given in Table \ref{tab:Regions}, in units of erg~s$^{-1}$. For the purpose of our analysis, these are converted to luminosities in solar units by dividing $L_{8.0}$ for the Sun's luminosity of $3.826\times 10^{33}~\mathrm{erg~s^{-1}}$.

Measurement errors are dominated by uncertainties in the centering of the photometric apertures on sources in the IRAC images. The errors were  estimated 
by `wiggling' the photometric aperture around seven representative regions by $\pm$3 pixels (about one resolution element in the 8~$\mu$m image) around the nominal centers and re-measuring photometry at both 3.6~$\mu$m and 8~$\mu$m. The resulting flux variations are then combined to propagate the uncertainty to the dust-only L$_{8.0}$. This uncertainty, at the level of 0.17~dex, is adopted as our measurement uncertainty for all sources. 

\section{Description of Models} \label{sec:models}

We employ the models produced by \citet{Lin} to compare our measurements with expectations for the dust emission of young star clusters. These authors combined the Starburst99 models \citep{Leitherer, Vazquez+2005} for the stellar populations with the \citet{DraineLi} models for the dust emission. We give below a short description of the  assumptions used by \citet{Lin} for both sets of models, and the quantities these authors derived for comparison with measurements and which will be used for the present study, as well.

Spectral energy distributions (SEDs) of stellar population models were generated by \citet{Lin} for both instantaneous burst and constant star formation models in the age range 1--500~Myr, with a metallicity range $Z=0.004-0.02$ (with $Z_{\odot}=0.014$), and covering several stellar evolutionary tracks. Instantaneous bursts are considered reasonable representations of the stellar populations contained  in star clusters \citep{Wofford+2016}. The model SEDs were then integrated in the wavelength range 0.0912--2~$\mu$m to produce the expected dust emission: $L(\mathrm{TIR})=f_{\mathrm{abs}}\times L(0.0912-2)$, where $L(\mathrm{TIR})$ is the infrared emission in the range 3--1000~$\mu$m, $L(0.0912-2)$ is the stellar light in the UV-to-nearIR range, and $f_{\mathrm{abs}}$ gives the fraction of the stellar light that is absorbed by the dust and re-emitted in the infrared. $f_{\mathrm{abs}}=1$ means that 100\% of the UV-to-nearIR stellar light from a star cluster is absorbed by dust; this is clearly a strict upper limit and unlikely to be generally true for our clusters, since they are all detected at optical wavelengths \citep{Sarajedini}.

The \citet{DraineLi} models were used to calculate the fraction of $L(\mathrm{TIR})$ emitted at 8 $\mu$m. The shape of the dust SED depends on various parameters, including the grain size distribution, the PAH abundance ($q_{\mathrm{PAH}}$), and the intensity of the starlight heating the dust. For the latter, \citet{DraineLi} use the dimensionless parameter $U$, which is the intensity of the local radiation field normalized to the radiation field intensity of the solar neighborhood. \citet{Lin} calculated a set of $L_{8.0}/L(\mathrm{TIR})$ ratios assuming the grain size distribution of the Milky Way, a fixed value for $U=12$, and a range of PAH abundances from $q_{\mathrm{PAH}}=0.47\%$ to 4.58\%. The latter value, $4.58\%$, is the highest available from the models and is consistent with the value observed in our own Milky Way. 

\citet{Lin} produced trends of the mass-normalized $L_{8.0}$ of a star cluster as a function of the cluster's age. The dependence of $L_{8.0}$ on stellar mass is a trivial one: lighter clusters contain proportionally less stars, thus less stellar light that can be processed by dust into the infrared. Normalizing $L_{8.0}$ by the cluster's mass removes this trivial dependence. The dependence on age is, conversely, more complex. For instantaneous burst SEDs, \citet{Lin} find that the PAH emission peaks at an age of $\approx$3~Myr and then decreases quickly from there. This is because young stellar populations have higher UV emission than older ones on account of the presence of massive stars; the sharp decrease beyond 3~Myr coincides with the most massive stars beginning to die off. The decreasing UV emission for ages $>$3~Myr causes a decrease in the amount of infrared emission that can  be produced by the UV-heated dust and a corresponding decrease in the  emission at 8~$\mu$m.  For the constant star formation models, the 8~$\mu$m  emission is enhanced relative to the instantaneous models at ages$\geq$5~Myr. However, like for the instantaneous model, the mass-normalized 8~$\mu$m luminosity in the constant star formation model shows a steady decrease with increasing age. This is because, although massive, UV bright stars are continuously produced in this model, the stellar mass accumulates with time, thus steadily increasing the denominator of the $L_{8.0}/M$ ratio.

After cluster age and mass, the next set of major dependencies for $L_{8.0}$ are given by the PAH abundance ($q_{\mathrm{PAH}}$) and the fraction of stellar light absorbed by dust ($f_{\mathrm{abs}}$). These two parameters, which vary independently, only impact the scaling of the models for $L_{8.0}/M$ as a function of cluster age,  but not their shape. \citet{Lin} determines that $L_{8.0}/L(\mathrm{TIR})$ is weakly sensitive on $U$ for the range of values expected in star-forming regions, and that changes in the properties of the stellar population models, e.g., metallicity and stellar evolutionary tracks, have negligible impact on the 8~$\mu$m emission. 

We consider Milky Way-like dust properties (Milky Way dust grain size distribution, $q_{\mathrm{PAH}}=4.58\%$) appropriate for M33, because the oxygen abundances of the regions where the star clusters are located are within the range $12+\log(\mathrm{O/H})\sim8.4-8.75$ ($\sim$0.5--1.2 solar)\footnote{We adopt an oxygen abundance of $12+\log(\mathrm{O/H})=8.69$ for the Sun \citep{Asplund+2009}.}, as measured from young stars and the gas \citep{U+2009, Bresolin+2011, Toribio+2016, Lin+2017}, and the PAH fraction is known to depend on metallicity \citep{Aniano+2020}. 

\section{Results} \label{sec:analysis}

\begin{figure}
\plotone{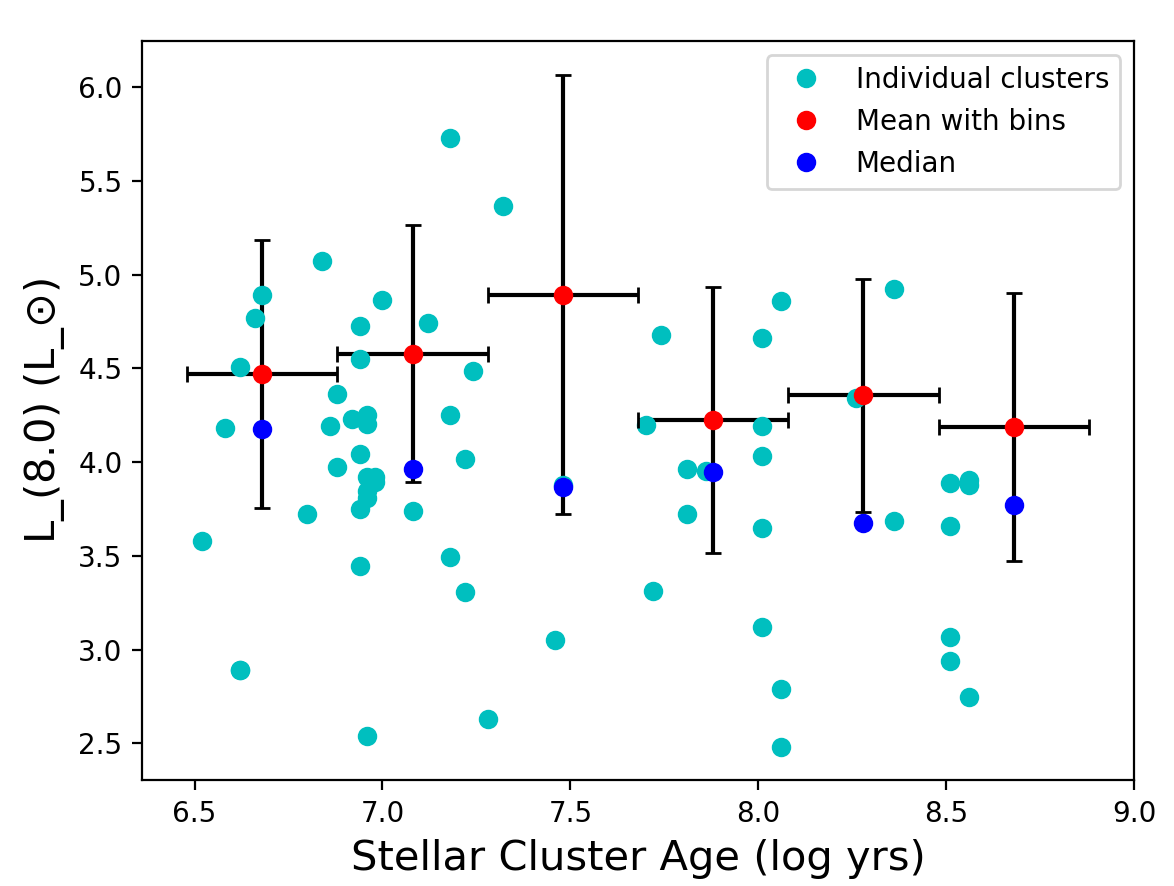}
\caption{The dust-only 8~$\mu$m luminosity of each cluster as a function of age. The mean (red points, with the 1~$\sigma$ standard deviation as black vertical lines) and the median (blue points) are shown for binned data, divided in six equal width bins in $\log(\mathrm{age})$. The size of the bins is shown with horizontal black bars. The bins' central values begin at $\log(\mathrm{age/yr}) = 6.68$ and increase in steps of 0.4~dex until the highest--age bin at $\log(\mathrm{age/yr})=8.68$.
\label{fig: original}}
\end{figure}

Prior to considering the mass-normalized 8~$\mu$m emission, we analyze the luminosity, $L_{8.0}$, itself as a function of cluster age. This is shown in Figure~\ref{fig: original}, where we find no correlation between these two quantities, as expected from the discussion in the previous section. The figure shows the individual star clusters as  well as both the mean and median of the data binned in six groups of age. Given the large scatter in the original data, which is reflected in a large standard deviation in the binned data, there is no relevant trend observed for the 8~$\mu$m luminosity also in the binned data. This confirms the argument presented in the previous section that the amount of dust emission from a star cluster depends not only on the presence of UV-emitting (young) stars, but also on the number of these stars, which is proportional to the total mass of the cluster if the stellar initial mass function is the same across all clusters.

\begin{figure}
\plotone{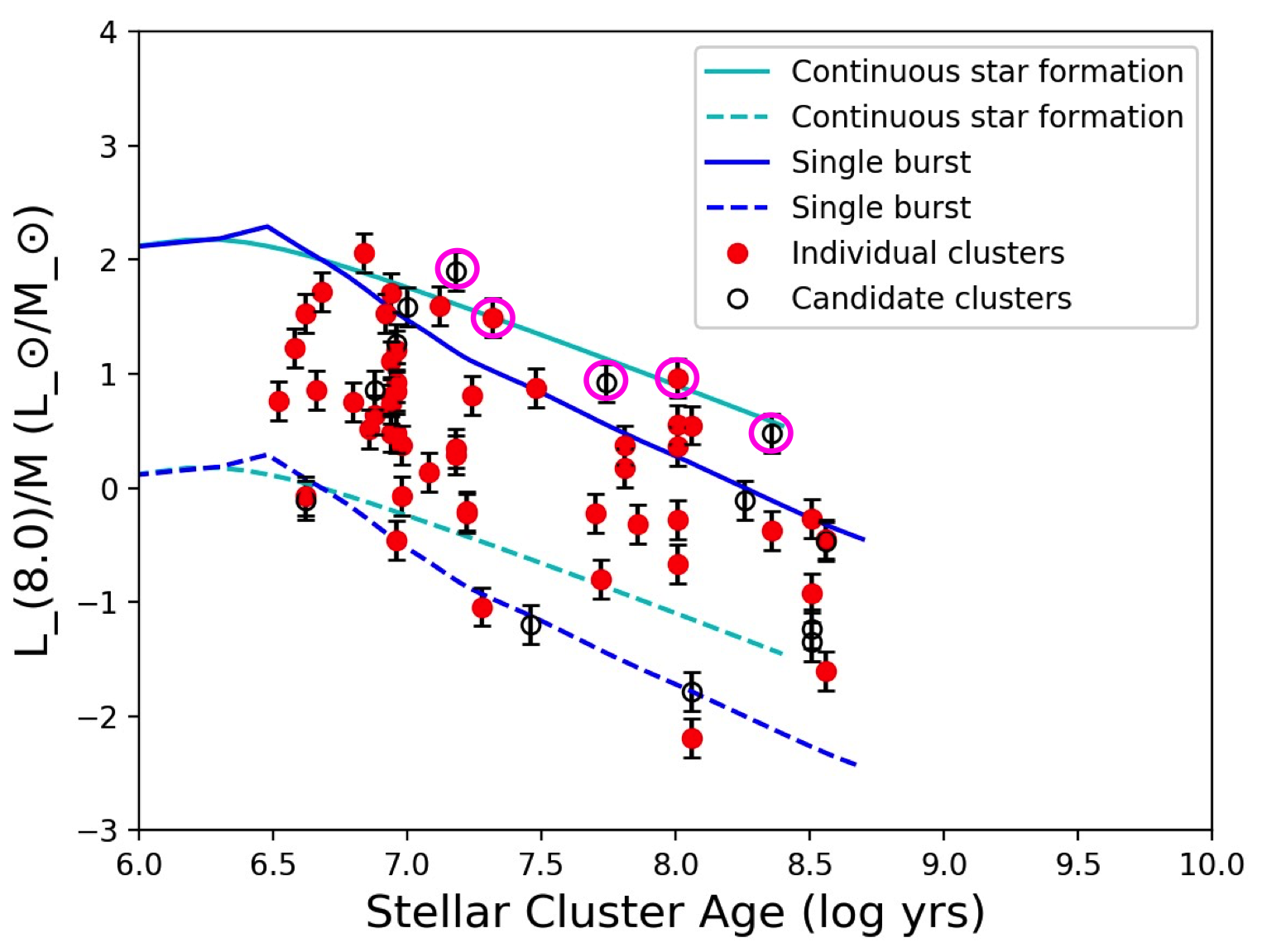}
\caption{The dust-only 8~$\mu$m luminosity divided by the stellar mass, $L_{8.0}/M$, as a function of age for both clusters (red points) and `unknowns' (black circles), shown together with their 1~$\sigma$ uncertainties (black bars). The model expectations are shown for both the instantaneous, single burst (blue lines) and the constant star formation case (cyan lines). The top-most lines show the models' location for the case $f_{\mathrm{abs}}=1$, i.e., that 100\% of the cluster's UV/optical light is absorbed by dust and re-emitted in the infrared. The lower lines show the location for $f_{\mathrm{abs}}=0.01$, i.e., only 1\% of the UV/optical light is absorbed by dust. These models bracket the observations. Magenta circles identify the five clusters/unknowns whose values of $L_{8.0}/M$ lie 2~$\sigma$ or more above the prediction for the  $f_{\mathrm{abs}}=1$ single burst model.} 
\label{fig: MWPower1_2}
\end{figure}

Next, we turn to the mass-normalized 8~$\mu$m emission, $L_{8.0}/M$, plotted as a function of the cluster's age; this is shown in Figure~\ref{fig: MWPower1_2}, where both clusters and the few sources classified as `unknown' in \citet{Sarajedini} are plotted. The mass-nomalized 8~$\mu$m luminosity displays a clear inverse correlation for increasing cluster's age. We overlay the models of \citet{Lin}, discussed in Section~\ref{sec:models}, on the data. The top-most model lines correspond to the case of instantaneous burst and continuous star formation populations with 100\% of the UV-to-nearIR luminosity absorbed by dust and re-emitted in the infrared. These models mark the upper envelope to the data, similar to what observed by \citet{Lin} for their data. The observed trend has an immediate interpretation in light of the models. For instantaneous burst models, the younger clusters have more massive stars, implying larger amounts of UV emission to heat the dust, while the older clusters lose their massive stars so the dust is heated less effectively and the infrared (and 8~$\mu$m) emission per unit stellar mass decreases. For constant star formation models, younger clusters have lower masses and, thus, proportionally higher $L_{8.0}/M$ than older clusters. Lower luminosity clusters at fixed age can be obtained by decreasing the fraction of UV luminosity absorbed by dust, which rigidly shifts the models downward. The lower end of the envelope is marked by model lines where only $\sim$1\% of the UV/optical luminosity is processed by dust into the infrared (dashed lines in Figure~\ref{fig: MWPower1_2}). 

\section{Discussion} \label{sec:discussion}

Our main result is that the mass-normalized 8~$\mu$m luminosity in star clusters and cluster candidates  (the `unknown' sources) sharply decreases for increasing age of the cluster population (Figure~\ref{fig: MWPower1_2}), thus confirming the results of \citet{Lin}. Over a factor 125 in age increase, from $\sim$3~Myr to 400~Myr, the normalized 8~$\mu$m luminosity decreases by about  two orders of magnitude.  At fixed age, our data show a significant a scatter, also of  about two orders of  magnitude. When compared with models for the expected dust emission of simple stellar populations, the data mostly fall below the $f_{\mathrm{abs}} = 1$ model lines, as expected since the models mark the upper envelope to the 8~$\mu$m luminosity. These models are derived under the assumption that 100\% of the UV/optical luminosity generated by the stars in the clusters is absorbed by dust and re-emitted in the infrared. This is clearly an extreme behavior, as we expect (and it is observed) most star clusters to emit at least a fraction of their UV emission directly, without dust absorption. The observed scatter at fixed age is in line with  this expectation; in Figure~\ref{fig: MWPower1_2} we show the location of the same models, but under the condition that only 1\% (lowest, dashed, model lines) of the UV/optical emission from the star clusters is absorbed by dust and re-emitted in the infrared. These two extreme model assumptions bracket the location of the vast majority ($\sim$90\%) of the data points within the uncertainties, indicating that the models are effective at explaining the observed behavior.

Although young star clusters are well represented by single stellar populations \citep{Wofford+2016}, we note that some of our 8~$\mu$m luminosity estimates exceed the predictions of the instantaneous burst models for ages$>$10$^7$~yr and at high dust absorption fractions (blue solid line in Figure~\ref{fig: MWPower1_2}). 
There are   5 clusters (8\% of total, magenta circles in Figure~\ref{fig: MWPower1_2}) whose mass-normalized 8~$\mu$m luminosity exceeds the topmost model prediction for instantaneous burst population by at least 2~$\sigma$ and are better explained by constant star formation (cyan solid line in Figure~\ref{fig: MWPower1_2}). \citet{Lin} find a similar discrepancy between the strongest dust emitting clusters and  instantaneous models in their sample. However, because of the distance of their targets (3.5--5~Mpc), \citet{Lin} cannot  exclude that part of the observed excess may be due to blends of star clusters with different ages or contamination from the background population, which could artificially increase the observed 8~$\mu$m luminosity relative to the assigned cluster age. This type of contamination is not applicable to our homogeneous sample from a single very nearby galaxy where blends of clusters can be easily recognized and removed. Furthermore, the emission from our star clusters is measured at $\sim$20~pc resolution, which removes much of the potential contamination from the background populations. If the four outliers were the result of constant star formation within the photometric aperture, the SFR of each region  would be  $\lesssim$10$^{-6}$~M$_{\odot}$~yr$^{-1}$, sufficiently low that star formation would be sporadic \citep{Weisz+2012}. We can therefore exclude constant star formation as an explanation for the five outliers.

We visually inspect the five outliers to garner additional insights into potential causes for their excess 8~$\mu$m luminosity. These outliers are two clusters (ID \# 215 and 250) and {three unknowns (ID \# 107, 276 and 336) from Table~\ref{tab:Regions}. Source \# 276 is shown in Figure~\ref{fig:Example}; it is located near another, brighter source whose 8~$\mu$m emission wings spill over into the photometric aperture used for the source's measurement. This morphology is representative of that presented by the other outliers as well. The proximity to brighter sources is likely to contribute to the observed excess in all five outliers.

There can be additional reasons for the excess dust emission of some star clusters relative to instantaneous model expectations. First and foremost, we do not have uncertainties for the ages and masses of our star clusters. While mass uncertainties for clusters in M33 are around several tens of percent, uncertainties in age determinations can be larger, in the range 0.1--0.3~dex \citep{Moeller+2021}. Adding such uncertainty to the ages, i.e., along the horizontal axis, would significantly reduce the discrepancy between data and instantaneous models. In addition, the ages of some of the clusters, which were derived from U-to-K band spectral energy distributions \citep[e.g.,][]{Ma+2001}, may be overestimated, since age determinations are uncertain in the absence of either UV or line emission information \citep{Whitmore2020}. Furthermore, there may be uncertainties in the PAH models, which are still being updated \citep{Draine2020}. Finally, star clusters may contain multiple-age stellar populations, generated over timescales of about 100~Myr \citep{Bastian+2018}, which may affect age determinations from single-age models like Starburst99.  

Our results expand over those of \citet{Lin} in two ways. Thanks to the proximity of M33, we expand by about an order of magnitude the luminosity range at fixed age towards fainter clusters. We do this across all ages, from 3~Myr to $\sim$400~Myr, which enables us to establish that, at fixed age, the mass-normalized 8~$\mu$m emission covers a range of at least two orders of magnitude at all the ages probed here. We also increase  about 5--fold (24 sources versus 5) the number of clusters in the age range 10$^7$--10$^8$~yr, thus increasing the statistics in this age range and reinforcing the results of \citet{Lin}. 

\citet{Lin} attributes the range of mass-normalized 8~$\mu$m emission at fixed age to two effects: (1) variations in $f_{\mathrm{abs}}$, the fraction of stellar UV/optical light absorbed by dust; and (2) variations in the PAH fraction of the dust. In the present case, we can exclude that variations in PAH fraction play a significant role in the observed scatter of $L_{8.0}/M$ at constant age. As discussed in Section~\ref{sec:models}, the oxygen abundance range at the location of our cluster sample is $\sim$0.5--1.2~solar; adopting the dependence of the PAH fraction on oxygen abundance from \citet{Aniano+2020}, this range translates into a factor 2--3 variation in PAH abundance and a similar variation in 8~$\mu$m luminosity \citep{Lin}. This is far below the two-orders-of-magnitude scatter of $L_{8.0}/M$ at constant age observed in our data. The large  scatter is more easily explained with variations in $f_{\mathrm{abs}}$ by a similar magnitude, from $f_{\mathrm{abs}}\sim 1$ to $f_{\mathrm{abs}}\sim 0.01$. 

\section{Conclusions} \label{sec:conclusions}

The main result of this analysis is that the 8~$\mu$m emission from dust displays a strong dependence on the age of the stellar population at young ages, decreasing by a factor $\sim$100 in mass-normalized intensity going from 3~Myr to $\sim$400~Myr . The decrease tracks expectations from models of emission from stars and dust, for a range of fractions of dust-absorbed UV/optical luminosity, from $\sim$100\% down to $\sim$1\%. This result can have important consequences for the detectability of galaxies at high redshift, where stellar populations tend to be young, although the dust fractions and PAH abundances tend to be low, lower than those found in a nearby galaxy like M33. Our study 
indicates that  the 8~$\mu$m luminosity can be effectively used as a tracer of the age of young stellar populations, although 
additional dependencies (metallicity, dust content, dust absorption fraction of the UV light, contamination from heating by older stellar populations, e.g.,  \citealt[][]{Calzetti+2007}) continue to complicate its use as a SFR indicator. In conclusion, we suggest that the  8~$\mu$m dust emission should not be used as a SFR indicator for individual star-forming regions or small galactic regions, where stochastic effects in the star formation history can dominate. Comparisons between galaxies or between galaxy populations will need to take into account potential differences in their star formation histories.

\begin{acknowledgments}
The authors thank  the anonymous referee for comments that  have greatly improved the manuscript.

Based on archival data from the {\sl Spitzer Space Telescope}, which was operated by the 
Jet Propulsion Laboratory, California Institute of Technology under a contract with NASA, and retrieved from the NASA/IPAC Infrared Science Archive, which is 
operated by the Jet Propulsion Laboratory, California Institute of Technology,  under  contract  with NASA. 

This research has also made use of the NASA/IPAC Extragalactic Database (NED) which is operated by the Jet
Propulsion Laboratory, California Institute of Technology, under contract with the National Aeronautics and Space
Administration.

Z. L. acknowledges support from the China Postdoctoral Science Foundation (2021M700137).
\end{acknowledgments}

\vspace{5mm}
\facilities{Spitzer Space Telescope (IRAC)}
\software{Astroconda (IRAF), SAOImage DS9, Python}

\startlongtable
\begin{deluxetable}{cccccccc}
\tablecolumns{8}
\tablenum{1}
\tablecaption{\label{tab:Regions}}
\tablewidth{0pt}
\tablehead{
\colhead{Region$^a$} & \colhead{RA, DEC$^a$} & \colhead{Log (Age)$^a$} & \colhead{Log (Mass)$^a$} & \colhead{$f_{3.6}^b$} & \colhead{$f_{8.0}^b$} & \colhead{$L_{8.0}$ $^c$} & \colhead{Classification$^a$}\\
\colhead{} & \colhead{(J2000)} & \colhead{(yrs)} & \colhead{(\(M_\odot\))} & \colhead{(MJy/sr)} & \colhead{(MJy/sr)} & \colhead{ergs/s} & \colhead{}
}
\decimalcolnumbers
\startdata
  42 & 01:33:10.11 30:29:56.9 & 6.58 & 2.95 & 143. & 72.8 & $5.91\times 10^{37}$ & Cluster \\ 
  48 & 01:33:14.29 30:27:11.1 & 6.96 & 2.99 & 13.7  & 50.4 & $6.99\times 10^{37}$ & Unknown \\
  53 & 01:33:16.10 30:20:56.0 & 6.96 & 3.00 & 8.29   & 2.82  & $1.36\times 10^{36}$ & Cluster \\
  54 & 01:33:16.63 30:34:35.7 & 8.01 & 3.48 & 21.0  & 33.3 & $4.21\times 10^{37}$ & Cluster \\
  68 & 01:33:22.32 30:40:59.4 & 8.36 & 4.06 & 8.30   & 14.7 & $1.90\times 10^{37}$ & Cluster \\
  70 & 01:33:23.10 30:33:00.5 & 7.86 & 4.27 & 20.5  & 28.2 & $3.47\times 10^{37}$ & Cluster \\
  74 & 01:33:23.90 30:40:26.0 & 6.92 & 2.70 & 46.3  & 55.3 & $6.61\times 10^{37}$ & Cluster \\
  80 & 01:33:26.00 30:36:24.3 & 8.51 & 4.42 & 26.6   & 9.18 & $4.55\times 10^{36}$ & Unknown \\
  85 & 01:33:26.75 30:33:21.4 & 7.08 & 3.60 & 24.1  & 20.0 & $2.14\times 10^{37}$ & Cluster \\
  86 & 01:33:26.94 30:34:52.6 & 6.94 & 2.96 & 27.1  & 13.5 & $1.08\times 10^{37}$ & Cluster \\
  105& 01:33:31.00 30:36:52.6 & 6.96 & 3.00 & 35.7  & 30.1 & $3.24\times 10^{37}$ & Cluster \\
  106& 01:33:31.10 30:33:45.5 & 7.81 & 3.55 & 23.0  & 19.2 & $2.05\times 10^{37}$ & Cluster \\
  107& 01:33:31.22 30:33:33.5 & 7.74 & 3.76 & 16.5 & 130. & $1.86\times 10^{38}$ & Unknown \\
  113& 01:33:32.01 30:33:21.8 & 6.98 & 3.52 & 77.6  & 38.4 & $3.05\times 10^{37}$ & Cluster \\
  114& 01:33:32.17 30:40:31.9 & 6.52 & 2.81 & 25.1  & 15.7 & $1.47\times 10^{37}$ & Cluster \\
  119& 01:33:32.72 30:36:55.2 & 6.84 & 3.01 & 43.0 & 321. & $4.61\times 10^{38}$ & Cluster \\
  141& 01:33:37.60 30:28:04.6 & 8.36 & 4.44 & 34.3 & 228. & $3.26\times 10^{38}$ & Unknown \\ 
  144& 01:33:38.04 30:33:05.4 & 7.22 & 3.51 & 27.5  & 11.6 & $7.86\times 10^{36}$ & Cluster \\
  152& 01:33:39.69 30:31:09.2 & 7.46 & 4.25 & 23.2  & 8.31  & $4.39\times 10^{36}$ & Unknown \\
  153& 01:33:39.71 30:32:29.2 & 7.12 & 3.14 & 61.7 & 159. & $2.14\times 10^{38}$ & Cluster \\
  172& 01:33:43.85 30:32:10.4 & 8.06 & 4.31 & 42.0 & 200. & $2.82\times 10^{38}$ & Cluster \\ 
  176& 01:33:44.51 30:37:52.7 & 8.26 & 4.45 & 34.1  & 65.5 & $8.54\times 10^{37}$ & Unknown \\ 
  181& 01:33:45.80 30:27:17.3 & 7.18 & 3.20 & 14.9  & 11.6 & $1.21\times 10^{37}$ & Cluster \\
  194& 01:33:50.70 30:58:50.3 & 7.28 & 3.67 & 7.28   & 2.79  & $1.66\times 10^{36}$ & Cluster \\
  195& 01:33:50.73 30:44:56.2 & 7.81 & 3.59 & 20.0  & 28.8 & $3.58\times 10^{37}$ & Cluster \\ 
  197& 01:33:50.85 30:38:34.5 & 7.22 & 4.24 & 29.3  & 34.0 & $4.03\times 10^{37}$ & Cluster \\
  198& 01:33:50.90 30:38:55.5 & 6.86 & 3.68 & 110. & 66.4 & $6.08\times 10^{37}$ & Cluster \\
  201& 01:33:51.24 30:34:13.2 & 6.96 & 3.00 & 14.8  & 45.7 & $6.25\times 10^{37}$ & Cluster \\
  214& 01:33:53.69 30:48:21.5 & 7.72 & 4.11 & 15.4   & 8.9 & $7.99\times 10^{36}$ & Cluster \\ 
  215& 01:33:54.10 30:33:09.7 & 7.32 & 3.87 & 49.4 & 621. & $9.03\times 10^{38}$ & Cluster \\
  217& 01:33:54.63 30:34:48.3 & 6.94 & 2.84 & 47.9 & 104. & $1.38\times 10^{38}$ & Cluster \\
  219& 01:33:54.75 30:45:28.4 & 6.62 & 2.98 & 2.92   & 85.6 & $1.26\times 10^{38}$ & Cluster \\  
  222& 01:33:55.18 30:47:58.0 & 7.70 & 4.42 & 28.1  & 47.9 & $6.12\times 10^{37}$ & Cluster \\ 
  228& 01:33:56.18 30:38:39.8 & 6.94 & 3.24 & 44.3  & 39.2 & $4.30\times 10^{37}$ & Cluster \\
  229& 01:33:56.21 30:45:51.8 & 8.01 & 3.83 & 18.0  & 45.3 & $6.09\times 10^{37}$ & Cluster \\ 
  243& 01:33:57.87 30:33:25.7 & 7.18 & 3.90 & 68.3  & 62.5 & $6.92\times 10^{37}$ & Cluster \\
  246& 01:33:58.03 30:39:26.2 & 7.24 & 3.68 & 43.9  & 90.9 & $1.20\times 10^{38}$ & Cluster \\
  250& 01:33:58.86 30:34:43.2 & 8.01 & 3.70 & 43.3 & 131. & $1.79\times 10^{38}$ & Cluster \\ 
  255& 01:33:59.52 30:45:49.9 & 6.88 & 3.72 & 38.8  & 69.6 & $8.98\times 10^{37}$ & Cluster \\ 
  257& 01:33:59.74 30:41:24.4 & 6.68 & 3.17 & 3.83 & 206. & $3.04\times 10^{38}$ & Cluster \\
  260& 01:34:00.01 30:33:54.3 & 6.98 & 3.99 & 146. & 55.6 & $3.26\times 10^{37}$ & Cluster \\
  269& 01:34:01.60 30:42:31.1 & 6.62 & 3.00 & 12.7  & 4.98  & $3.05\times 10^{36}$ & Unknown \\
  271& 01:34:01.75 30:32:25.7 & 8.51 & 4.16 & 22.7  & 25.5 & $3.00\times 10^{37}$ & Cluster \\
  272& 01:34:01.99 30:38:10.9 & 6.62 & 2.96 & 12.5  & 4.93  & $3.03\times 10^{35}$ & Cluster \\
  276& 01:34:02.48 30:38:41.1 & 7.18 & 3.83 & 90.2 & 1427. & $2.08\times 10^{39}$ & Unknown \\
  280& 01:34:02.79 30:46:36.8 & 7.00 & 3.28 & 35.7 & 201 & $2.85\times 10^{38}$ & Unknown \\
  288& 01:34:03.83 30:29:33.5 & 8.51 & 4.18 & 23.4   & 7.68 & $3.40\times 10^{36}$ & Unknown \\
  310& 01:34:07.28 30:38:29.5 & 8.01 & 3.79 & 13.0   & 6.44 & $5.12\times 10^{36}$ & Cluster \\
  320& 01:34:08.53 30:39:02.4 & 6.66 & 3.91 & 45.3 & 164. & $2.27\times 10^{38}$ & Cluster \\
  325& 01:34:08.96 30:36:33.8 & 6.88 & 3.12 & 66.4  & 40.1 & $3.67\times 10^{37}$ & Unknown \\
  329& 01:34:10.09 30:45:29.4 & 6.96 & 3.33 & 38.1  & 25.8 & $2.52\times 10^{37}$ & Cluster \\
  335& 01:34:11.35 30:41:27.9 & 8.56 & 4.35 & 10.1  & 23.3 & $3.11\times 10^{37}$ & Cluster \\
  336& 01:34:11.36 30:41:27.9 & 8.56 & 4.35 & 9.63   & 22.1 & $2.94\times 10^{37}$ & Unknown \\
  347& 01:34:14.02 30:39:29.5 & 8.01 & 3.93 & 14.8  & 15.2 & $1.74\times 10^{37}$ & Cluster \\ 
  351& 01:34:14.65 30:32:35.0 & 8.56 & 4.35 & 11.6   & 4.2 & $2.18\times 10^{36}$ & Cluster \\
  355& 01:34:15.51 30:42:11.5 & 8.51 & 4.58 & 44.5  & 22.2 & $1.77\times 10^{37}$ & Cluster \\  
  361& 01:34:17.54 30:42:36.7 & 7.48 & 3.00 & 7.51   & 21.6 & $2.95\times 10^{37}$ & Cluster \\
  370& 01:34:19.44 30:46:21.2 & 8.06 & 4.57 & 29.2 & 8.33   & $2.39\times 10^{36}$ & Unknown \\
  372& 01:34:20.17 30:39:33.3 & 6.96 & 3.00 & 9.29   & 20.6 & $2.73\times 10^{37}$ & Cluster \\
  410& 01:34:33.09 30:37:36.3 & 6.80 & 2.97 & 9.13   & 16.0 & $2.05\times 10^{37}$ & Cluster \\
  420& 01:34:40.41 30:46:01.3 & 6.94 & 3.99 & 220. & 190. & $2.07\times 10^{38}$ & Cluster \\
  422& 01:34:40.72 30:53:02.0 & 6.94 & 2.63 & 31.3  & 21.9 & $2.18\times 10^{37}$ & Cluster \\
  439& 01:34:50.10 30:47:04.1 & 8.06 & 4.67 & 29.9 & 7.68   & $1.18\times 10^{36}$ & Cluster \\
\enddata
\tablecomments{$^a$ IDs, coordinates, ages, masses and classifications from \citet{Sarajedini}. Masses are rescaled to our adopted distance of 850~kpc for M33, which  is slightly smaller than the one used by \citet{Sarajedini}.}
\tablecomments{$^b$ Flux density measured in the 5$^{\prime\prime}$ radius photometric aperture, uncorrected for any aperture effect.}
\tablecomments{$^c$ Calculated for the adopted distance of 850 kpc for M33. The values listed include all corrections described in Sec~\ref{sec:processing}.}
\end{deluxetable}

\bibliographystyle{aasjournal}
\bibliography{references}{}

\end{document}